\documentclass{article}
\usepackage{arxiv}

\usepackage{amsmath}
\usepackage{makecell}
\usepackage[utf8]{inputenc} 
\usepackage[T1]{fontenc}    
\usepackage{hyperref}       
\usepackage{url}            
\usepackage{booktabs}       
\usepackage{amsfonts}       
\usepackage{nicefrac}       
\usepackage{microtype}      
\usepackage{lipsum}
\usepackage{cite}
\usepackage{graphicx}
\usepackage{xcolor}
\usepackage{placeins}
\usepackage[group-separator={,}]{siunitx}
\graphicspath{ {./figs/} }

\makeatletter
\renewcommand*{\@fnsymbol}[1]{\ensuremath{\ifcase#1\or \,1\or \,2\or \,3\or
   \mathsection\or \mathparagraph\or \|\or **\or \dagger\dagger
   \or \ddagger\ddagger \else\@ctrerr\fi}}
\makeatother
   
\title{Accelerating Phase Field Simulations Through a Hybrid Adaptive Fourier Neural Operator with U-Net Backbone}

\author{Christophe Bonneville\thanks{Sandia National Laboratories, Livermore, CA 94550; \texttt{\{cpbonne,ahegde,hnnajm,csafta\}@sandia.gov}} 
\and \textbf{Nathan Bieberdorf}\thanks{Materials Sciences Division, Lawrence Berkeley National Laboratory \texttt{and} Department of Materials Science and Engineering, University of California, Berkeley, CA 94720, USA; \texttt{\{nbieberdorf,mdasta\}@berkeley.edu}} 
\and  \textbf{Arun Hegde}\footnotemark[1]
\and  \textbf{Mark Asta}\footnotemark[2]
\and  \textbf{Habib N. Najm}\footnotemark[1]
\and  \textbf{Laurent Capolungo}\thanks{Los Alamos National Laboratory, Los Alamos, NM 87544; \texttt{laurent@lanl.gov}}
\and  \textbf{Cosmin Safta}\footnotemark[1]
}

\begin{document}

\maketitle
\begin{abstract}
Prolonged contact between a corrosive liquid and metal alloys can cause progressive dealloying. For one such process as liquid-metal dealloying (LMD), phase field models have been developed to understand the mechanisms leading to complex morphologies. However, the LMD governing equations in these models often involve coupled non-linear partial differential equations (PDE), which are challenging to solve numerically. In particular, numerical stiffness in the PDEs requires an extremely refined time step size (on the order of $10^{-12}s$ or smaller). This computational bottleneck is especially problematic when running LMD simulation until a late time horizon is required. This motivates the development of surrogate models capable of leaping forward in time, by skipping several consecutive time steps at-once. In this paper, we propose \textit{U-Shaped Adaptive Fourier Neural Operators} (U-AFNO), a machine learning (ML) based model inspired by recent advances in neural operator learning. U-AFNO employs U-Nets for extracting and reconstructing local features within the physical fields, and passes the latent space through a vision transformer (ViT) implemented in the Fourier space (AFNO). We use U-AFNOs to learn the dynamics mapping the field at a current time step into a later time step. We also identify global quantities of interest (QoI) describing the corrosion process (e.g. the deformation of the liquid-metal interface, lost metal, etc.) and show that our proposed U-AFNO model is able to accurately predict the field dynamics, in-spite of the chaotic nature of LMD. In particular, our model reproduces the key micro-structure statistics and QoIs with a level of accuracy on-par with the high-fidelity numerical solver. Finally, we also investigate the opportunity of using hybrid simulations, in which we alternate forward leap in time using the U-AFNO with high-fidelity time stepping. We demonstrate that while advantageous for some surrogate model design choices, our proposed U-AFNO model in fully auto-regressive settings consistently outperforms hybrid schemes. 
\end{abstract}


\section{Introduction}
\label{sec:intro}

Phase field modeling is a powerful tool for simulating microstructure and morphological evolution of materials systems subjected to various driving forces (e.g. chemical, mechanical, electrical, etc.). The phase field method relies on a  “diffuse-interface” approach to tracking phases, interfaces, as well as conserved quantities (e.g. mole-fraction of an element). This allows for temporal evolution of the system to be defined everywhere by a set of partial differential equations (PDEs) representing well-known continuum and interfacial kinetic processes \cite{LQChenReview,KarmaReview,SteinbachReview}. While this avoids the numerical complexity associated with explicitly tracking and updating a sharp interface in addition to providing a route to describing interface reactions, phase field modeling can still require a high-resolution spatial discretization to accurately resolve nano-scale interfacial phenomena. This can render large length- and time-scale phase field simulations prohibitively computationally-expensive, thereby limiting the use of phase field modeling to coarse grain microstructure evolution.

Several approaches have been developed to circumvent the computational expense of phase field simulations. One approach is to use adaptive meshing to set the spatial discretization to be very fine near the interface and much coarser in the bulk phases where longer-range transport processes take place \cite{Provatas1999}. However, this is not guaranteed to be advantageous in situations where the interface morphology is rapidly changing, since frequent re-initialization of the adaptive mesh would be required. Another approach 
is to artificially broaden interfaces – from the nanometer-scale to, for example, the micron-scale – and thereby resolve the system with a relatively coarse mesh at interfaces as well as within bulk phases. In this approach, transport equations and effective kinetic parameters are rederived to ensure the phase field simulations evolve consistently with a realistic thin-interface system \cite{KarmaThinInterface1996,KarmaThinInterface1998,KarmaAntiTrapping}. But any approach using a micron-scale interface is inherently limited in its ability to resolve morphological features that initiate at the nano-scale. And so, while these approaches are valuable for a wide variety of phase field simulations, they are not necessarily suitable for situations with fine-scale, dynamic morphologies. 

One such example is the dealloying corrosion of metals, where the infiltration of a corrosive agent into the alloy can rapidly lead to morphologically-complex metal structures \cite{Harrison1959,Erlebacher2001,McCue_Karma_Erlebacher_2018,MCCUE201610,Wada:2023,Liu2021}. Understanding corrosion-induced dealloying is particularly critical in energetic applications, such as next-generation molten-salt nuclear reactor design \cite{CALDERONI2012842, ROPER2022108924}. In these aforementioned metal structures, preferential dissolution of the alloy precedes an instability that initiates nano-scale, compositionally-heterogenous features along the interface \cite{Erlebacher2001}. In a coupled manner, these features grow while the corrosive agent infiltrates as individual channels into the alloy. Reemergence of the nano-scale interfacial instability can cause these channels to bifurcate and advance tortuously into the alloy, increasing the alloy’s topological complexity \cite{geslin2015}. Simultaneously, capillary forces act to coarsen the dealloyed structure and reduce local curvatures and overall topological complexity \cite{ChenWiegart2012}. Competition among these processes sets up a dynamic morphology that underlies the rate of corrosion into the base metal. Phase field simulations for dealloying corrosion, specifically for liquid metal dealloying, have successfully reproduced the experimentally-observed morphologies and uncovered the key mechanisms by which these structures evolve \cite{geslin2015,Lai:2022a,Lai:2022,bieberdorf2023}. To simulate representative domains and time-scales with high resolution, these phase field simulations have generally relied on using high performance computing resources and advanced time-integration schemes \cite{geslin2015,bieberdorf2023}. However, the length- and time-scales of such simulations are several orders of magnitude below experiments, so that validation requires extrapolation using scaling laws.

The computational cost of numerical simulations such as the ones involved in liquid-metal dealloying is a major bottleneck. This is particularly true when a large number of forward simulations are required, like in uncertainty quantification \cite{10754/656260,Smith2013UncertaintyQ}, inverse problems \cite{10754/656260,https://doi.org/10.1002/nme.2746,https://doi.org/10.1002/2016RS005998, peherstorfer2018survey}, and design optimization \cite{do1,do2, forrester2008engineering} applications. This computational challenge has historically motivated the development of surrogate models that are faster than the high-fidelity simulations, albeit being less accurate. This drop in accuracy can be an acceptable compromise when the time gain is significant. A particular class of surrogate models for PDEs, such as \textit{reduced-order-models} (ROMs) aim to decrease computational cost by reducing the dimensionality of the problem. This can be done by projecting snapshots of high fidelity simulations into a lower-dimensional latent space. Linear projection methods, such as proper-orthogonal decomposition (POD), are widely used for building accurate ROMs \cite{doi:10.1146/annurev.fl.25.010193.002543,rbm,benner2015survey, benner2017model, Safonov1988ASM}, and have been successfully applied to a variety of problems, e.g. in fluid mechanics~\cite{Stabile_2018,https://doi.org/10.1002/num.21835,Copeland_2022,https://doi.org/10.48550/arxiv.2201.07335,MCLAUGHLIN20162407, holmes2012turbulence}, structural dynamics~\cite{amsallem2009method, georgiou1999dynamics, guyan1965reduction}, and control systems~\cite{antoulas2005approximation, krener2008reduced, strazzullo2018model}. More recently, employing machine learning (ML) approaches to compress high fidelity data snapshots has gained significant interest. Indeed, non-linear mappings, such as neural networks, can yield better performance especially in advection-dominated problems \cite{LEE2020108973, kutz_2017, BONNEVILLE2024116535, bonneville2024comprehensive}. 

With recent advances in deep learning and computer vision, a second class of surrogate models for PDEs has emerged, namely \textit{neural operators} \cite{kovachki2021neural, li2021fourier, Lu_2021}. Neural operators learn mappings between function spaces through a resolution-invariant tensor-to-tensor regression. Popular neural operator architectures include Fourier neural operator (FNO) \cite{li2021fourier}, DeepONets \cite{Lu_2021}, and more recently Laplace Neural Operator (LNO) \cite{cao2023lno} and Convolutional Neural Operator (CNO) \cite{raonić2023convolutional} constructions. In time-dependent problems, neural operators can be employed to predict the physical state at a future time step, given the current state. Often, this strategy is used auto-regressively, i.e. the operator output becomes its own input, allowing for predicting consecutive time steps. This strategy (sometimes referred to as auto-regressive \textit{roll-out}) is analogous to conventional explicit time integrators. Auto-regressive neural operators roll-outs have the convenient advantage of being naturally well suited for non-parametric initial conditions (e.g. random initial field), unlike more traditional ROM approaches. As such, neural operators have been applied to a variety of problems, including fluid dynamics \cite{li2021fourier, a16010024, fno_les}, electromagnetics and electro-convection \cite{10.1007/978-3-031-36021-3_24, CAI2021110296}, fracture mechanics \cite{GOSWAMI2022114587}, weather prediction \cite{pathak2022fourcastnet,10.1145/3592979.3593412}, phase field modeling \cite{oommen_deeponet, wen2022ufno} and plasma physics \cite{gopakumar2023fourier}.

While neural operators have shown great accuracy in the aforementioned papers, recent work suggests that this accuracy can be increased further by combining key neural operator building blocks with U-Nets \cite{ronneberger2015unet} and Vision transformers (ViTs) \cite{dosovitskiy2021an}. Recent papers proposed to employ the encoder-decoder architecture of a U-Net as a feature extraction mechanism, in combination with a neural operator bottleneck \cite{wen2022ufno,diab2023udeeponet,10.1063/5.0158830}. Building on this work, ViTO \cite{ovadia2023vito} and DiTTO \cite{ovadia2023realtime} proposed a neural operator architecture based on ViTs to capture interdependence relations within the input field compressed features, where the compression is obtain by a U-Net. In this paper, we propose to build on these ideas by strategically placing an \textit{Adaptive} Fourier Neural Operator (AFNO) \cite{guibas2022adaptive} at the intermediate bottleneck of a U-Net. An AFNO is a special implementation of a ViT, where the attention mechanism specific to transformer architectures is computed in the Fourier space. This idea, initially introduced with FNO \cite{li2021fourier} allows for significant computational efficiency gains during training. We call our proposed model U-AFNO, and show that it is able to capture the phase field dynamics occurring in LMD with excellent accuracy, in-spite of the highly chaotic underlying physics. 

LMD phase fields are typically used to to extract local and global quantities of interest (QoIs). Such QoIs, describing, for example, the topology of corroded/dealloyed metal interfaces or the chemical composition within the alloy may be crucial for decision making, optimization, or uncertainty quantification purposes. In the present study, we employ several physically relevant QoIs to describe the dealloying process at both local and global scales. Furthermore, we show that our proposed U-AFNO model is accurate, not only according to state-of-the-art error metrics \cite{li2021fourier, Lu_2021, kovachki2021neural, li2022learning, oommen2023rethinking}, but also in correctly reproducing the aforementioned QoIs, unlike other commonly used neural operator models. Lastly, we also investigate the performance of blending our proposed surrogate model with high-fidelity simulations. In a recent ML-based phase field surrogate study \cite{oommen2023rethinking}, Oommen \textit{et al.} proposed to augment a U-Net-based model with alternate high-fidelity solver time stepping. After each leap in time through a U-Net forward pass, several of the following time steps are computed with the high-fidelity solver. This hybrid approach still partially relies on the high-fidelity solver, and is consequently significantly slower than fully auto-regressive roll-outs, but allows for greater (and tunable) accuracy. In this paper, we investigate the performance of U-AFNOs with and without hybrid high-fidelity time stepping. In particular, we show that even without hybrid time stepping augmentation, our proposed U-AFNO model can outperform other augmented models, such as U-Nets.

The contributions of this paper are therefore two-fold: (1) a novel surrogate model for LMD phase field simulation to accurately reproduce the micro-structure dynamics and the QoIs describing dealloying processes and (2) a thorough performance analysis of hybrid ML surrogates blended with high-fidelity solvers. Note that we employ the term \textit{high-fidelity} to refer to simulations obtained from traditional PDE solvers (sometimes referred to as \textit{full-order-models} in the literature), as opposed to surrogate models. In section~\ref{sec:lmd-hf}, we first discuss the mathematical model for LMD processes and provide brief details of our high-fidelity solver implementation. Then, in section~\ref{sec:uafno}, we introduce our proposed U-AFNO surrogate model, and, in section~\ref{sec:QOI}, we discuss the various local and global QoIs employed in the paper. Finally, in section~\ref{sec:app}, we present and discuss the performance of our proposed model and summarize this work in section~\ref{sec:concl}.

\section{High-Fidelity Phase Field Model for Liquid-Metal Dealloying}
\label{sec:lmd-hf}

Consider a model dealloying system, where a binary alloy composed of two species, denoted as `A’ and `B’, is in contact with a liquid dealloying agent of pure species `C’ (e.g. molten salt). The model is parameterized to represent a liquid metal dealloying system, following \cite{geslin2015,bieberdorf2023}. The solid and liquid phases are tracked using a non-conserved phase field variable $\phi$, where $\phi=1$ in the solid phase, $\phi=0$ in the liquid phase, and $0<\phi<1$ represents their diffuse interface. Conserved phase field variables are used to track each species, with $c_\text{A}$, $c_\text{B}$, and $c_\text{C}$ representing the mole-fraction of species A, B, and C, respectively. Everywhere in the system, $c_\text{A}+c_\text{B}+c_\text{C}=1$.

\begin{figure}[!ht]
\centering
    \includegraphics[width=1\textwidth]{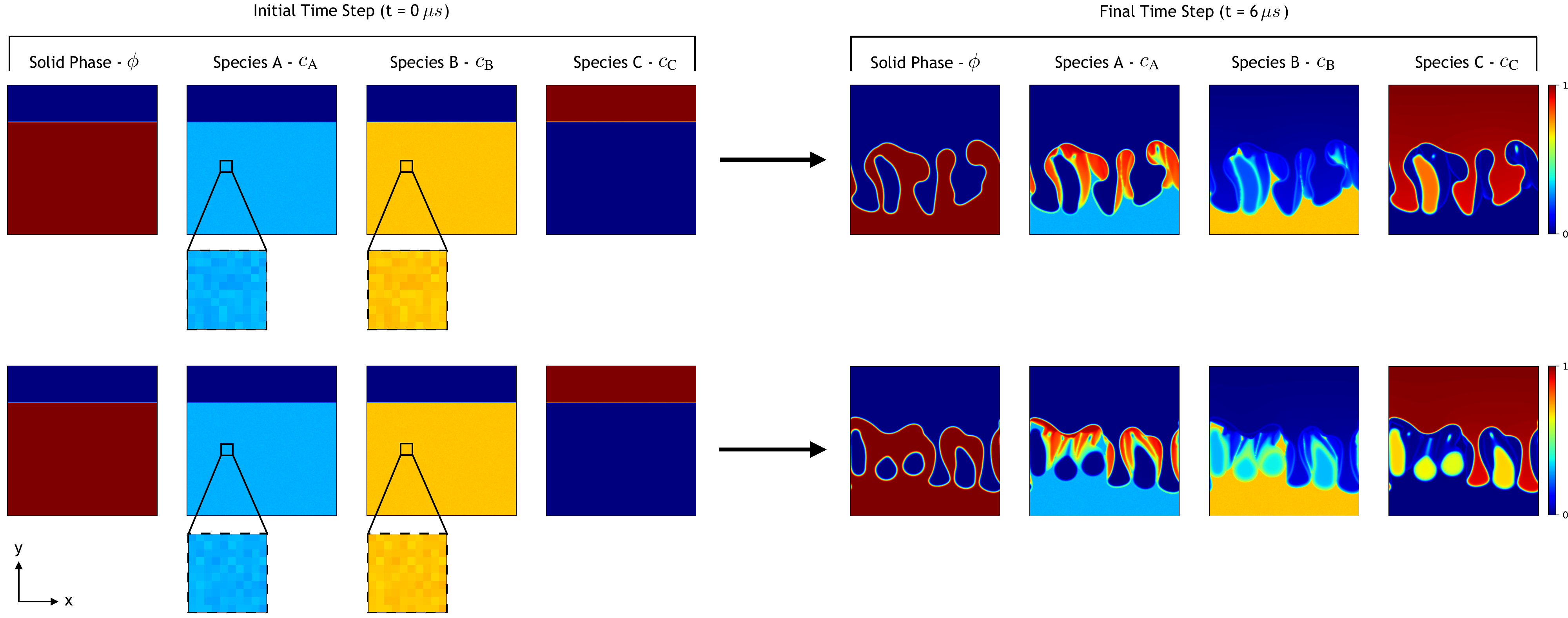}
    \caption{Liquid metal dealloying with two sets of initial conditions. The initial species fields ($c_A$ and $c_B$ at $t=0$) are contaminated with low-amplitude random white noise (left fields). Given the chaotic nature of the dealloying process, the initial noise perturbation eventually leads to widely different solid phase fields at late time (e.g. right fields, at $t=6\,\mu$s, after 6 million time steps).}
    \label{fig:initialCondition}
\end{figure}

While the system in principle exhibits dynamics in 3-dimensional (3D) space, we focus here on a 2D context for computational convenience, presuming uniformity in the third dimension. The initial condition for the system is shown in figure \ref{fig:initialCondition}. The spatial domain $\Omega$ is defined as a square of width 102.4 nm ($\Omega=[0, 102.4]\times[0, 102.4]$). The binary alloy is initialized in the lower part of the domain, while the liquid phase is on top. The alloy initial composition is nominally $(c_\text{A},c_\text{B})=(0.3,0.7)$ everywhere, with superimposed white noise sampled from a uniform distribution in the interval $[-0.025, 0.025]$ at each grid-point. The liquid is initialized as pure $c_\text{C}=1$. Dirichlet boundary conditions of $\phi=0$ and $c_\text{C}=1$ are enforced along the top edge of the simulation. Neumann boundary conditions (zero normal gradient) are enforced along the bottom edge of the simulation. Periodic boundary conditions are enforced on the left and right edges of the domain. We discretize the system using a regular 512$\times$512 grid, with a grid spacing of $\Delta x=0.2$~nm.


The evolution of the phase field is defined using
\begin{equation}
\label{eq:phase}
    \frac{\partial \phi}{\partial t}=-\tilde{M}_\phi \frac{\pi^2}{8 \eta} \frac{\delta F}{\delta \phi}
\end{equation}
where $\tilde{M}_\phi$ is the mobility of the solid-liquid interface, $\eta$ is the diffuse interface width, and ${\delta F}/{\delta \phi}$ is the functional derivative of the free energy functional, $F$, with respect to the solid phase~\cite{SteinbachReview, Steinbach1999}. Conserved-species mole fraction evolution is governed by the continuity equation
\begin{equation}
\label{eq:specie}
    \frac{\partial c_i}{\partial t}=\nabla\cdot \sum_{j=\text{A,B}}{ M_{ij}(\phi)\nabla\bigg(\frac{\delta F}{\delta c_j}\bigg) } 
\end{equation}
with a phase-dependent solute mobility $M_{ij}$ and the functional derivative ${\delta F}/{\delta c_j}$. Eq.~\eqref{eq:specie} is solved for species A and B evolution, while species C can always be determined by the condition $c_\text{C}=1-c_\text{A}-c_\text{B}$. The summation in Eq.~\eqref{eq:specie} is only accomplished for species A and B since ${\delta F}/{\delta c_j}$ represents the diffusion potential of species $j$ with respect to species C. The solute mobility is given by
\begin{equation}
    \label{eq:soluteMobilitiy}
    M_{ij}(\phi) = D_\text{liq}(1-\phi)\frac{V_a}{kT}c_i(\delta_{ij}-c_j)
\end{equation}
where the diffusivity is assumed to be negligibly small in the solid phase and smoothly increases to $D_\text{liq}$ in the liquid phase. $V_\text{a}$ is an atomic volume, $k$ is the Boltzmann constant, and $T$ is temperature. 

The free energy functional is defined as
\begin{equation}
    F=\int_\Omega (f_\text{phase}(\phi)+f_\text{chem}(\phi,c_\text{A},c_\text{B}) ) dV
\end{equation}
with the free energy density $f_\text{phase}$ that sets the diffuse interface, defined as
\begin{equation}
    f_\text{phase}=\frac{4\sigma_\text{sl}}{\eta}\bigg[ \frac{\eta^2}{\pi^2}(\nabla\phi)^2 + \phi(1-\phi) \bigg]
\end{equation}
where $\sigma_\text{sl}$ is the interfacial energy of a pure solid-liquid interface. The chemical energy density $f_\text{chem}$ is given by a regular solution model, which has an ideal entropy of mixing and non-ideal enthalpy of mixing.
\begin{equation}
    \label{eq:fchem}
    f_\text{chem} = \sum_{i=\text{A,B,C}} \bigg[ \frac{kT}{V_\text{a}} c_i \ln(c_i) \bigg] + \Omega_\text{AC}c_\text{A}c_\text{C} + h(\phi)\Delta g_\text{sl} + \frac{1}{2}\sum_{i=\text{A,B,C}} \big[ \kappa(\nabla c_i)^2 \big]
\end{equation}
where $\Omega_\text{AC}$ is the excess enthalpy of mixing between species A and C, $h(\phi)$ is an interpolation function, $\Delta g_\text{sl}$ is the difference between solid and liquid reference chemical energies, and $\kappa$ is the energy penalty for spatial gradients of each species. The interpolation function, defined to smoothly vary between $h(\phi=1)=1$ in the solid and $h(\phi=0)=0$ in the liquid, is given by
\begin{equation}
    h(\phi) = \frac{1}{2} + \frac{2}{\pi} \bigg[(2\phi-1)\sqrt{\phi(1-\phi)} + \frac{1}{2}\arcsin{(2\phi-1)} \bigg]
\end{equation}
while $\Delta g_\text{sl}$ is defined as
\begin{equation}
    \Delta g_\text{sl} = \sum_{i=\text{A,B,C}}\bigg[ c_i L_i \bigg(\frac{T-T_i}{T_i}\bigg) \bigg]
\end{equation}
where $L_i$ is the latent heat of melting and $T_i$ is the melting temperature for each species $i$.

The parameters for the model are given in Table~\ref{tab:PFparameters}. In this work, species A is assumed to have a very small solubility with the liquid C bath, which is set by the large, positive excess enthalpy of mixing $\Omega_\text{AC}\gg0$ and the large melting temperature of species A, $T_\text{A}\gg T$. This leads to the dealloying condition where species B is selectively dissolved from the alloy into the liquid C bath, while A rearranges laterally along the interface to form a topologically-complex solid structure. Importantly, diffusion along the interface (where $0<\phi<1$) is significantly fast, according to Eq.~\eqref{eq:soluteMobilitiy}, to enable reorganization of species A. We assume the dealloying is rate-limited by diffusion in the liquid phase (based on~\cite{MCCUE201610,geslin2015}), and therefore set the interface mobility $\tilde{M}_\phi$ to be very large so that dissolution is fast compared to diffusion in the liquid.

\begin{table}[ht]
\centering
\begin{tabular}{clccl}
\Xhline{3\arrayrulewidth}\\[-6pt]
Parameter & Value &   & Parameter & Value\\[4pt]\hline\\[-6pt]
$T$ & 1775 K &   & $\sigma_{\text{sl}}$ & 0.2 J$\cdot$m$^{-2}$ \\
$\eta$ & 4 $\times 10^{-9} \text{m}$ &   & $\kappa$ & 2.4 $\times 10^{-9}$ J$\cdot$m$^{-1}$\\
$L_{\text{A}}$ & 2.82 $\times 10^{9}$ J$\cdot$m$^{-3}$ &   & $L_{\text{B}}$ & 1.89 $\times 10^{9}$  J$\cdot$m$^{-3}$\\
$L_{\text{C}}$ & 1.84 $\times 10^{9}$  J$\cdot$m$^{-3}$ &   & $T_{\text{A}}$ & 3290 K \\
$T_{\text{B}}$ & 1941 K &   & $T_{\text{C}}$ & 1358 K \\
$V_{\text{a}}$ & 0.01 $\times 10^{-27} \text{m}^{3}$ &   &$\Omega_{\text{AC}}$ & 1.44 $\times 10^{10}$ J$\cdot$m$^{-3}$\\
$\tilde{M}_{\phi}$ & 12.0 m$\cdot$s$^{-1}$$\cdot$GPa$^{-1}$ &   & $D_{\text{liq}}$ & $7\times 10^{-9}$ m$^2$$\cdot$s$^{-1}$\\[2pt]\Xhline{3\arrayrulewidth}
\end{tabular}
\vspace{0.05in}
\caption{Thermodynamic and kinetic parameters used in the high fidelity phase field simulations \cite{geslin2015,bieberdorf2023}}
\label{tab:PFparameters}
\end{table}

In the high-fidelity phase field simulations, non-conserved phases are updated using a simple forward Euler time-integration of Eq.~\eqref{eq:phase}. However, for conserved species, Eq.~\eqref{eq:specie} is numerically very stiff, and using forward Euler would require prohibitively small time-steps to maintain numerical stability. This is, in part due to fine spatial-resolution of the numerical mesh and also the fourth-derivative on each species field in Eq.~\eqref{eq:specie} that arises from the Laplacian in Eq.~\eqref{eq:fchem}. Instead of using forward Euler, conserved species are updated using a semi-implicit time-integration of Eq.~\eqref{eq:specie}, which is based on the approach of~\cite{Badalassi2003} and uses spectral methods. The reader is referred to~\cite{bieberdorf2023} for details of this semi-implicit time-integration scheme and implementation. These high fidelity phase field simulations use a time-step size of $\Delta t=10^{-12}$ s.

\section{Leaping forward in time with U-AFNO}
\label{sec:uafno}
In this section, we introduce our proposed U-AFNO model. We first cover attention mechanisms and Fourier neural operators (section \ref{sec:afno}), and further motivate the need for a U-Net backbone augmentation (section \ref{sec:uafno-subsec}). Finally, we also briefly introduce hybrid high-fidelity and surrogate schemes (section \ref{sec:hybrid}).
\subsection{Adaptive Fourier Neural Operator}
\label{sec:afno}
Vision transformers (ViT)~\cite{dosovitskiy2021an} have become widely popular in computer vision~\cite{Han_2023, Khan_2022}, and recent attempts to use them for operator learning have shown promising early results~\cite{hao2024dpot,VARGHESE2024106086,li2023transformer,ovadia2023vito}. Originally introduced for natural language processing applications~\cite{vaswani2023attention}, the key idea of transformers is to compute a similarity metric (\textit{attention}) across sub-samples of the training data (\textit{tokenisation}). In ViTs, this is done by splitting the input image (denoted $X$) into $N$ smaller patches $x\in\mathbb{R}^d$ (each patch is reformatted into a vector of dimension $d$). The input image can thus be represented by a tensor $X\in\mathbb{R}^{N\times d}$. The attention can be computed with the following expression:
\begin{equation}
\label{att}
    Att:\mathbb{R}^{N\times d}\mapsto\mathbb{R}^{N\times d}\hspace{0.5in}Att(X)=\sigma\bigg(\frac{XW_q(XW_k)^\top}{\sqrt{d}}\bigg)XW_v
\end{equation}
Where $\sigma$ is an activation function, generally taken as softmax. The query, key and value matrices, respectively $W_q$, $W_k$ and $W_v$ are learnable weights. Since ViT are designed for 2D image inputs, Eq.~\eqref{att} complexity is quadratic in nature. Thus, the self-attention mechanism of ViTs scales poorly with the patch size and the resolution of the input image (or equivalently in this paper, the input physical field). This is particularly problematic for two reasons. First, high resolution images of physical fields are sometimes necessary to capture small-scale local features. Second, a small patch size is desirable as it generally yields higher prediction accuracy~\cite{beyer2023flexivit}. As a result, training ViTs can be computationally challenging. To alleviate this issue, Adaptive Fourier Neural Operators (AFNOs)~\cite{guibas2022adaptive} have been proposed. The main idea of AFNOs is to formalize Eq.~\eqref{att} as a kernel summation. Each row of $Att(X)$ can be re-written as:
\begin{equation}
    Att(X)(s,:)=\sum_{t=1}^NX(t,:)\cdot\underbrace{K(s,t) W_v}_{\kappa(s,t)}\hspace{0.5in}K=\sigma\bigg(\frac{XW_q(XW_k)^\top}{\sqrt{d}}\bigg)
\end{equation}
Where $t\in[\![1,N]\!]$ represents a dummy row index (in this section only). By assuming that the kernel is a Green's kernel, i.e., $\kappa(s,t)=\kappa(s-t)$, self attention can be seen as a global convolution operation. Building on an idea originally introduced with Fourier Neural Operators (FNOs)~\cite{li2021fourier}, the attention operation is performed in the Fourier space, which allows reducing the convolution to a simpler and computationally faster element-wise matrix product:
\begin{equation}
\label{fft}
    Att(X)(s,:)\Longleftrightarrow\mathcal{F}^{-1}(\mathcal{F}(\kappa)\cdot\mathcal{F}(X))(s,:)
\end{equation}
Where $\mathcal{F}$ refers to the discrete FFT operation. Additional details on AFNOs can be found in~\cite{guibas2022adaptive}. It should be noted that unlike other popular neural operator models, such as FNOs~\cite{li2021fourier} and DeepONets~\cite{Lu_2021}, AFNOs are not neural operators in the true sense. Indeed, neural operators are generally considered to be resolution independent. While the operation described in Eq.~\eqref{fft} is resolution independent by construction, in AFNOs the input image is processed through an embedding layer before computing attention. This embedding layer is generally fully connected, with a fixed dimension, resulting in the loss of resolution invariance.

\subsection{U-AFNO: Adaptive Fourier Neural Operator with a U-Net Backbone}
\label{sec:uafno-subsec}

AFNOs have been successfully employed to build surrogates in a variety of physical simulation problems, such, as and most notably, weather forecasting through auto-regressive roll-out~\cite{pathak2022fourcastnet,10.1145/3592979.3593412}. However, AFNOs are prone to exhibit squared-shape artifacts and mismatch between patch interfaces in the predicted field. In our experiments, we have found this to be particularly true with the binary phase fields describing LMD simulations. To address this issue and ensure both meaningful feature extraction and smooth outputs, we propose to combine a simple AFNO with a U-Net backbone (U-AFNO). U-Nets~\cite{ronneberger2015unet} employ an encoder-decoder architecture, with U-shaped successive convolutions, downsampling, and upsampling layers, and skipped connections between the encoder and decoder. U-Nets have proven to be highly successful in image regression tasks, and combining the strengths of U-Nets with neural operators has been recently found to be useful~\cite{wen2022ufno,diab2023udeeponet,10.1063/5.0158830, oommen2023rethinking}. In particular, ViTO~\cite{ovadia2023vito} and DiTTO~\cite{ovadia2023realtime} studies have proposed to use vision transformer blocks within the standard U-Net bottleneck, along with a trunk network for enforcing resolution independence (as originally introduced with DeepONets~\cite{Lu_2021}). In the present study, we adopt a similar, yet simpler, approach by replacing the U-Net bottleneck with an AFNO. Note that downsampling the input image through the U-Net encoder has the convenient advantage of allowing for patch sizes of $1\times1$ (without downsampling, such small patch size would be computationally intractable).

\begin{figure}[!ht]
\centering
    \includegraphics[width=1\textwidth]{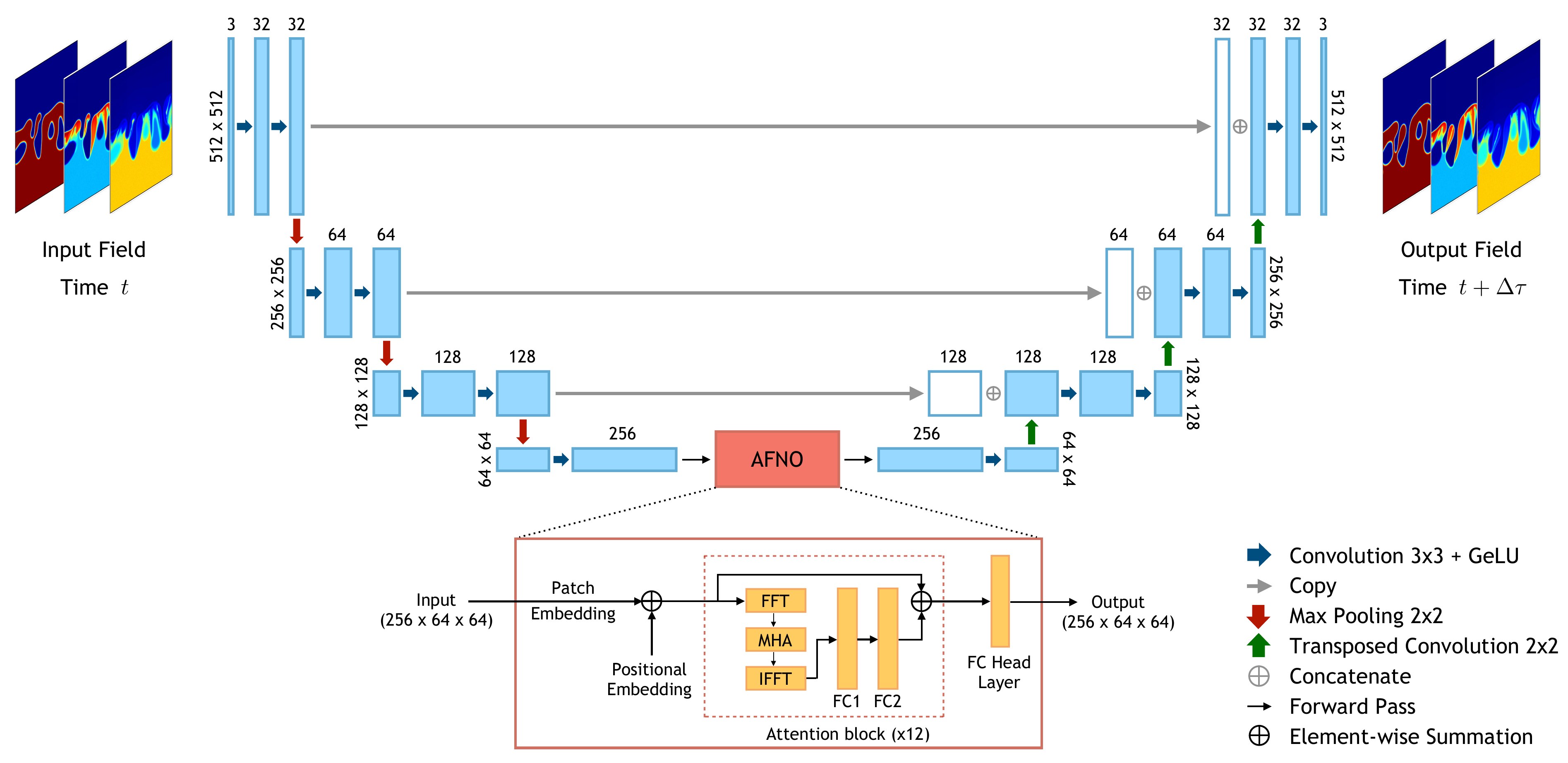}
    \caption{U-AFNO Architecture - The model takes as input the field at time $t$ and outputs the field at a future time step $t+\Delta\tau$. The U-Net color scheme is analogous to the one employed in the original U-Net paper~\cite{ronneberger2015unet}. The abbreviations in the AFNO layers are the following: FFT: Fast Fourier Transform, IFFT: Inverse Fast Fourier Transform, MHA: Mutli-Head Attention, FC: Fully Connected.}
    \label{fig:uafno}
\end{figure}

The architecture of the U-AFNO is represented in Figure~\ref{fig:uafno}. The input field has dimensions $3\times512\times512$ (the 3 channels correspond to $\phi$, $c_A$ and $c_B$ respectively -- note that $c_C$ is not needed here since its computation knowing $c_A$ and $c_B$ is trivial), and is downsampled to dimensions $256\times64\times64$ through the U-Net encoder. The AFNO bottleneck is composed of 12 back-to-back Attention blocks, each of which employs 16 attention heads, a patch size of $1\times1$, a skipped connection and two fully-connected layers with 3072 neurons each. The AFNO output is then upsampled back into the original dimension through the U-Net decoder. Both the U-Net and the AFNO layers use GeLU activation functions, and the final output of the U-AFNO is passed through a sigmoid (logistic) function. This guarantees that the U-AFNO predictions are bounded between 0 and 1, and hence consistent with the phase field physics.

\subsection{Hybrid High-Fidelity and Surrogate Model Coupled Solver}
\label{sec:hybrid}
Neural operators and adjacent regression models have been extensively used with an auto-regressive roll-out to predict PDE dynamics over time~\cite{li2021fourier, Lu_2021, kovachki2021neural, pathak2022fourcastnet}. While this approach is simple and fast, errors tend to grow quickly, potentially unbounded. Similarly to traditional explicit numerical solvers, auto-regressive roll-out may exhibit instabilities and/or un-physical behavior over time. This is in part because the outputs of neural operators are generally not guaranteed to lie within the correct manifold of physical fields (especially with fully data-driven operators). In order to slow-down error growth, a recently proposed approach~\cite{oommen2023rethinking, oommen_deeponet} alternates forward passes through the neural operator (forward leap in time) with high-fidelity time stepping (\textit{relaxation} steps). This strategy allows for greater flexibility in balancing the trade-off of speed-up vs. accuracy. More relaxation steps means greater reliance on the high-fidelity solver and fewer neural operator forward passes, which mitigates error build-up (as each new operator forward pass inevitably introduces some error), but also takes more time. Conversely, fewer relaxation steps means faster surrogate predictions but larger error accumulation in the long term. In this paper, we investigate the performance of U-AFNOs with and without these high-fidelity relaxation steps (in the former case, we employ the term \textit{hybrid simulation}),


\section{Phase Field Quantities of Interest}
\label{sec:QOI}
An essential motivation for developing physical simulations (whether high-fidelity or surrogate ones) is that it is often a necessary first step for extracting informative quantities of interest (QoIs) from the physical fields. These QoIs, typically computed locally and/or globally across the spatial field and over time may be crucial for decision making, optimization or uncertainty quantification purposes. In some applications, the extracted QoIs may be more important than the simulated physical field itself. In dealloying corrosion, some QoIs are particularly informative, such as the local curvature~\cite{McCue_Karma_Erlebacher_2018,Wada:2023,VAKILI20171852} and the total length (perimeter)~\cite{Tran_2019} of the solid-liquid interface, the penetration depth of the corrosive liquid into the metal~\cite{Lai:2022,MCCUE201610}, and the amount of remaining metal, species A and B, throughout the corrosion process. 

\subsection{Curvature and Perimeter}

\begin{figure}[!ht]
\centering
    \includegraphics[width=1\textwidth]{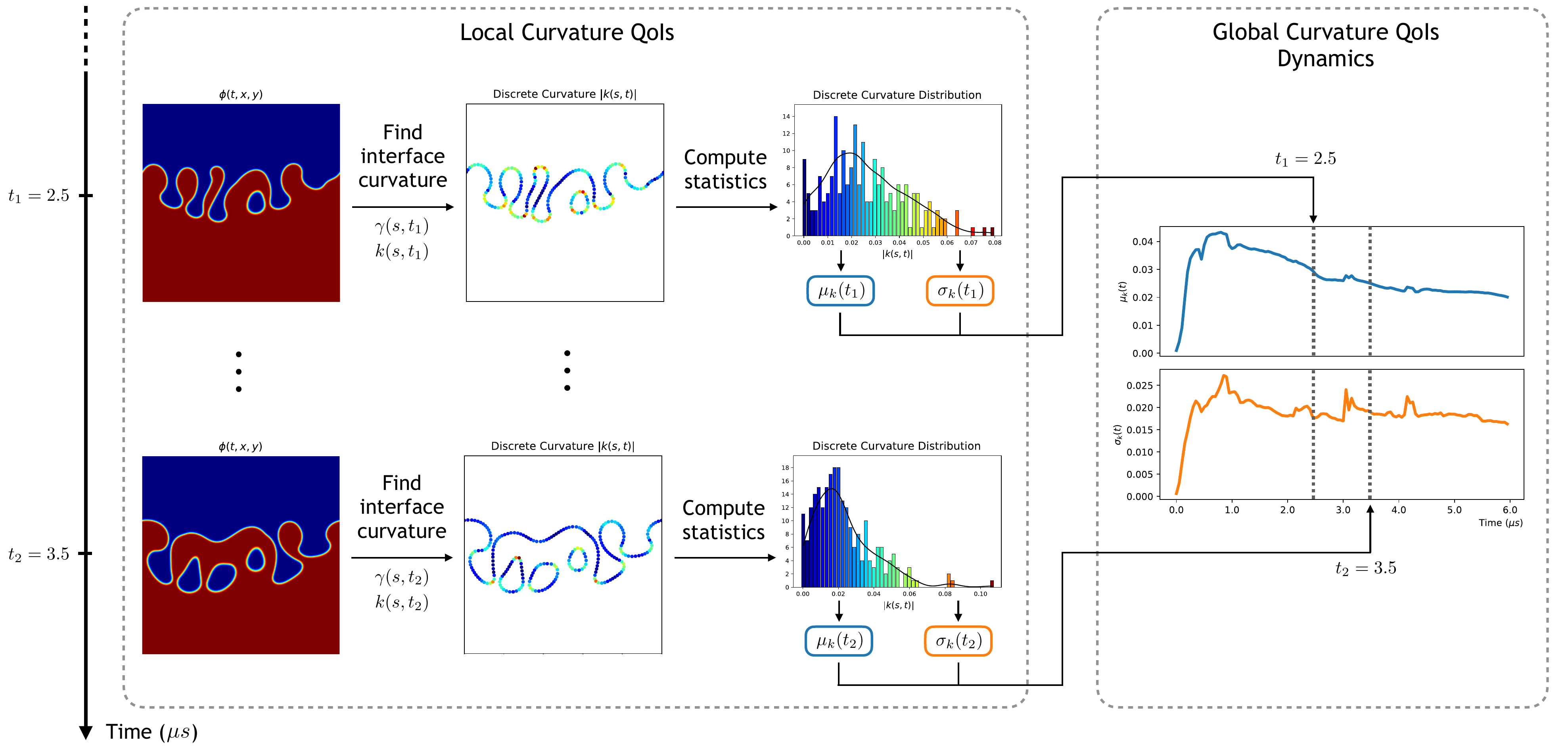}
    \caption{Curvature QoIs - At each time $t$, the parametric curve $\gamma(s,t)$ delineating the interface between the two phases is computed, from which we compute the curvature $k(s,t)$. Histograms of the curvature distribution at two different times ($t=2.5~\mu$s and $t=3.5~\mu$s) are shown, from which means and standard deviations are estimated and computed over time (blue and orange plots).}
    \label{fig:qoi}
\end{figure}

The first collection of QoIs we consider involve the curvature of the liquid-metal interface. We formally define the interface in this 2D system as any spatial coordinate $(x,y)\in\Omega$ such that $\phi(t,x,y)=0.5$, $\forall t$. The interface coordinates may then be described at any time $t$ as a parameterized curve, $\gamma(s,t)=(x(s,t),y(s,t))$ with $s\in[0,1]\subset\mathbb{R}$, and the signed curvature along the interface is:
\begin{equation}
\label{eq:curv}
    k(s,t)=\frac{x'y''-y'x''}{(x'^2+y'^2)^{3/2}}\hspace{0.3in}x'\equiv\frac{\partial x}{\partial s}\hspace{0.3in}y'\equiv\frac{\partial y}{\partial s}
\end{equation}
The curvature is defined locally, but we can quantify it globally by considering statistics such as the mean and standard deviation of the absolute curvature over the entire interface, at any moment in time:
\begin{equation}
\label{eq:curvms}
\mu_k(t)=\int_0^1|k(s,t)|\text{d}s\hspace{0.5in}\sigma_k(t)=\bigg[\int_0^1(|k(s,t)|-\mu_k(t))^2\text{d}s)\bigg]^{1/2}
\end{equation}
Furthermore, we also define the interface perimeter at any time as:
\begin{equation}
\label{eq:perim}
    p(t)=\int_0^1\sqrt{(x')^2+(y')^2}\text{d}s
\end{equation}
Eqs.~\eqref{eq:curv}, \eqref{eq:curvms} and \eqref{eq:perim} are all computed discretely using finite difference. Figure~\ref{fig:qoi} represents the general approach to compute the curvature locally at each time step, and extract statistical quantities to represent more global curvature dynamics over time. Note that in liquid metal dealloying, metal ligaments may fully separate from the rest of the solid phase and form islands. Similarly, some metal ligaments may rejoin and close-up, trapping liquid bubbles inside the metal. In some cases, these features can shrink via capillary forces until they are eventually eliminated. During such events, the mean and standard deviation of the curvature is naturally skewed upward, before sharply dropping once an island or bubble disappears. Such patterns can be observed in Figure~\ref{fig:qoi}.

\subsection{Penetration Depth}
The second collection of QoIs considered describe corrosive liquid penetration within the metal. As the species C-rich liquid progressively penetrates into the metal, the latter reorganizes to form narrow ligaments that typically grow as the liquid penetrates further into the bulk metal. We can quantify the average ligament length over time, as well as the maximum penetration depth, by considering the local extrema of the parametric curve representing the interface (along the vertical axis). The set of local minima and maxima at time $t$ are defined as:
\begin{equation}
    \mathcal{S}_\text{min}(t)=\{y(s,t)\,\,|\,\,y'=0\,\,\,\text{and}\,\,\,y''>0\}\hspace{0.5in}\mathcal{S}_\text{max}(t)=\{y(s,t)\,\,|\,\,y'=0\,\,\,\text{and}\,\,\,y''<0\}
\end{equation}
Each local maximum is representative of a ligament tip $y-$axis coordinate, while each local minimum is representative of a ligament base $y-$axis coordinate. Thus, we approximate the mean ligament height $\mu_d$ as the difference between the mean of the set of local minima and local maxima, thus
\begin{equation}
    \mu_d(t)=\mathbb{E}[\mathcal{S}_\text{max}(t)]-\mathbb{E}[\mathcal{S}_\text{min}(t)].
\end{equation}
We also define the maximum penetration depth as
\begin{equation}
    \text{max}_p(t)=1-\min(\mathcal{S}_\text{min}(t)).
\end{equation}

\subsection{Mass Loss}
The third and last set of QoIs considered describes the dynamics of lost material throughout the corrosion process. We define the total ``mass" of metal, species A, and species B as the integrals over the spatial domain of $\phi$, $c_A$ and $c_B$ respectively, thus
\begin{equation}
\label{eq:masses}
m_\phi(t)=\int_\Omega\phi(t,x,y)\text{d}x\text{d}y\hspace{0.5in}m_A(t)=\int_\Omega c_\text{A}(t,x,y)\text{d}x\text{d}y\hspace{0.5in}m_B(t)=\int_\Omega c_\text{B}(t,x,y)\text{d}x\text{d}y
\end{equation}
The actual masses in proper mass-units are related to these quantities with proportionality constants, these being the metal density in the case of $m_\phi$, and the molar weights for $m_\text{A}$ and $m_\text{B}$.

\section{Application}
\label{sec:app}
In this section, we present and discuss the performances of our U-AFNO model, both for predicting the full micro-structure fields as well as the QoIs described in the previous section. We also provide details about the model training and review the different error metrics used in this study. 

\subsection{Training}

The training set comprises 87 high-fidelity simulations, each containing 100 field snapshots spanning from time $t=1\mu$s to $t=6\mu$s. The U-AFNO is trained to leap \num{50000} (high-fidelity) time steps ahead, i.e. $\Delta\tau=5\cdot10^4\times\Delta t=0.05\mu$s every forward pass. The test set comprises 20 high-fidelity simulations of similar form. The model is trained for 20 epochs, using the Adam optimizer~\cite{kingma2017adam} with a learning rate of $10^{-4}$. The code is implemented with Pytorch~\cite{paszke2019pytorch} and the AFNO bottleneck is based on the FourCastNet implementation~\cite{pathak2022fourcastnet}. For baseline comparison, we also train a standard U-Net, an AFNO (with no U-Net backbone), and an FNO. All our models are trained on a single Nvidia A100 80GB GPU. The U-AFNO training took 35 hours.

In this paper, we run test surrogate simulations for up to time step $6\times 10^6$ (i.e. $t=6\mu$s), all starting with quasi-identical initial conditions. 
We initialize the surrogate simulations with time integration of the high-fidelity model (in the following sections, $n_\text{init}=\num{1000000}$ steps). This allows for the instabilities at the interface between the phases to grow and be sufficiently distinguishable across different initial conditions. We test both \textit{fully auto-regressive} simulations (i.e. high-fidelity relaxation steps are not employed, except during initialization), and hybrid simulations with different numbers of high-fidelity relaxation steps (as presented in section \ref{sec:hybrid}).

\subsection{Error Metrics}

We consider several error metrics to assess the accuracy of our model. While the relative error between the ground truth (i.e. high-fidelity simulation) and the predicted field is a common error metric in the neural operator literature~\cite{li2021fourier, Lu_2021, kovachki2021neural}, it is not directly applicable here due to the chaotic nature of LMD simulations. In chaotic systems, comparing invariant statistical properties, such as field auto-correlations is more meaningful and informative~\cite{li2022learning}. In phase field simulations, the field spatial auto-correlation is equivalent to the probability that two points in the field will be part of the same phase~\cite{HERMAN2020589}. As such, the spatial auto-correlation relative error has been shown to be more relevant for describing the micro-structure statistics than other relative error metrics, and has been recently used in neural-operator-based phase field models~\cite{oommen2023rethinking}. The spatial auto-correlation relative error is defined as:
\begin{equation}
\label{releleq}
    e_\text{AC}(\mathbf{u}, \mathbf{\hat{u}},t)=\frac{|\!|S_{\mathbf{\hat{u}}\mathbf{\hat{u}}}(\pmb{r},t)-S_{\mathbf{u}\mathbf{u}}(\pmb{r},t)|\!|_2}{|\!|S_{\mathbf{u}\mathbf{u}}(\pmb{r},t)|\!|_2}
\end{equation}
where $\mathbf{u}$ can represent either a ground truth species mole fraction or phase field variable at time $t$, and $\mathbf{\hat{u}}$ represents the corresponding surrogate prediction. $S$ is the corresponding spatial auto-correlation function and $\pmb{r}$ is a vector connecting any two points in the spatial field. Note that, for baseline comparison, we also compute the average auto-correlation relative \textit{discrepancy} between each pair of ground truth simulation contained in the test set. In this later case, equation \ref{releleq} is used as well, but we employ the term \textit{discrepancy} rather than \textit{error} since we only compare ground truth simulations together. This discrepancy metric allows for meaningful judgement of the difference between ground truth dynamics and our surrogate, compared to a high-fidelity simulation with a different random initial condition. To assess the ability of our model to predict the key QoIs described in section \ref{sec:QOI}, we also consider a second set of error metrics, the relative error of each QoI across all time steps. For example, we define the mean curvature relative error as:
\begin{equation}
e(\pmb{\mu_k},\pmb{\hat{\mu}_k})=\frac{|\!|\pmb{\hat{\mu}_k}-\pmb{\mu_k}|\!|_2}{|\!|\pmb{\mu_k}|\!|_2}
\end{equation}
where $\pmb{\mu_k}$ is the vector of ground truth mean curvature at each time step: $\pmb{\mu_k}=(\mu_k(t_0),\dots,\mu_k(t_\text{max}))$ (and $\pmb{\hat{\mu}_k}$ is the equivalent vector for the surrogate prediction). In a similar fashion, we define the relative error for the curvature standard deviation, the interface perimeter, the maximum penetration depth, the mean ligament height, and the remaining metal mass.

\subsection{Results}

\FloatBarrier

\begin{figure}[!ht]
\centering
    \includegraphics[width=1\textwidth]{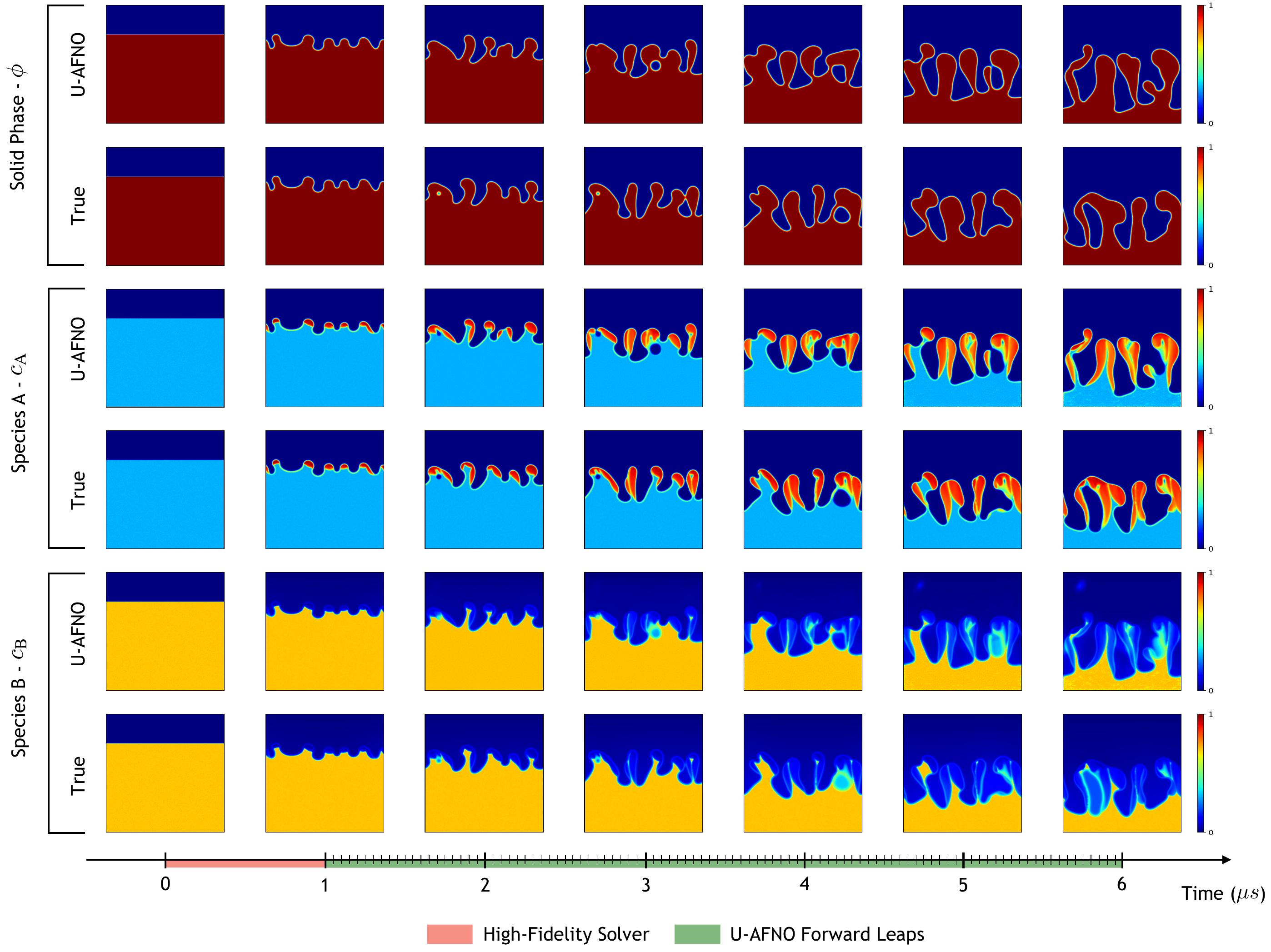}
    \caption{U-AFNO-B/1 field predictions in the \textbf{fully auto-regressive case}. The predicted and ground truth fields are shown at $t=(0, 1, 2, 3, 4, 5, 6)~\mu$s. Since the initial $10^6$ time steps are computed with the high-fidelity solver, time $t=0-1~\mu$s for the U-AFNO and the ground truth are identical.}
    \label{fig:field_far}
\end{figure}

\begin{figure}[!ht]
\centering
    \includegraphics[width=1\textwidth]{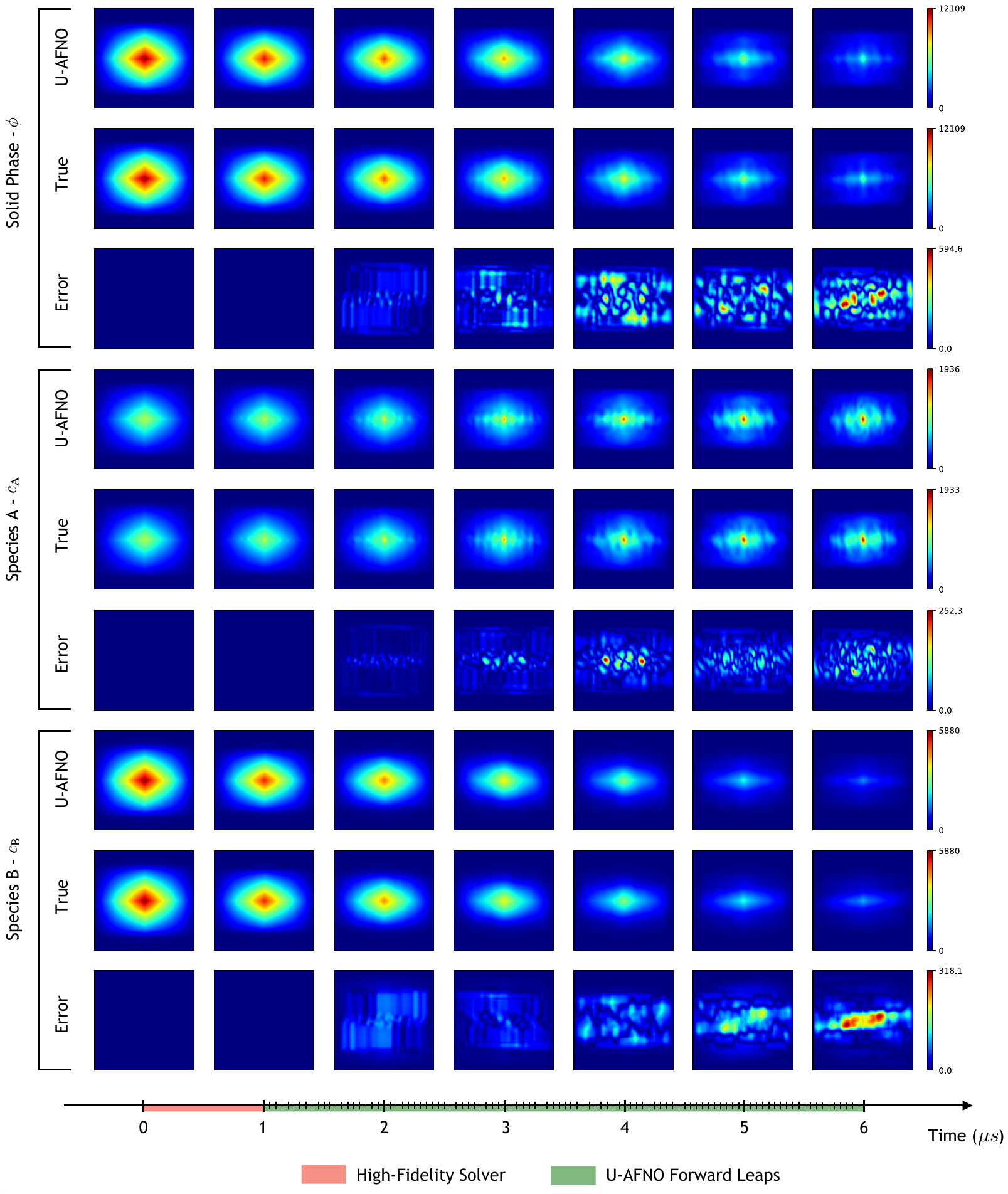}
    \caption{U-AFNO field auto-correlation maps in the \textbf{fully auto-regressive case}. The auto-correlation maps correspond to the fields shown in figure \ref{fig:field_far}. The error maps are the absolute error between the predicted and ground truth auto-correlation.}
    \label{fig:field_autocor}
\end{figure}

We first consider the fully auto-regressive case. Figure~\ref{fig:field_far} shows the solid phase $\phi$ and species A and B predicted with a U-AFNO-B/1 surrogate (where B/1 refers to a $1\times1$ patch size), at different moments in time from $t=0\,\mu$s to $t=6\,\mu$s. The total number of auto-regressive forward passes through the U-AFNO is $100$. Due to the chaotic behavior of the LMD physics, the surrogate model is not able to reproduce the exact ligament shapes available from the ground truth data, but it captures patterns that are visually consistent with them. Specifically, based on visual inspection, the ligaments of the solid phase exhibit satisfactory thickness, height and size. Further, the liquid phase penetrates into the metal in a manner consistent with the ground truth results, making its way steadily towards the bottom boundary. The dynamics of both species are similarly consistent; especially at the solid-liquid interface where species A tends to segregate to the growing ligaments while species B is dissolved into the liquid. This behavior, consistent with dealloying phenomena, is correctly reproduced here. This empirical observation can be confirmed by looking at the associated auto-correlation maps, represented in Figure~\ref{fig:field_autocor}. The predicted auto-correlations closely match the ground truth, indicating that the U-AFNO reproduces invariant field statistics well.

\begin{figure}[!ht]
\centering
    \includegraphics[width=0.85\textwidth]{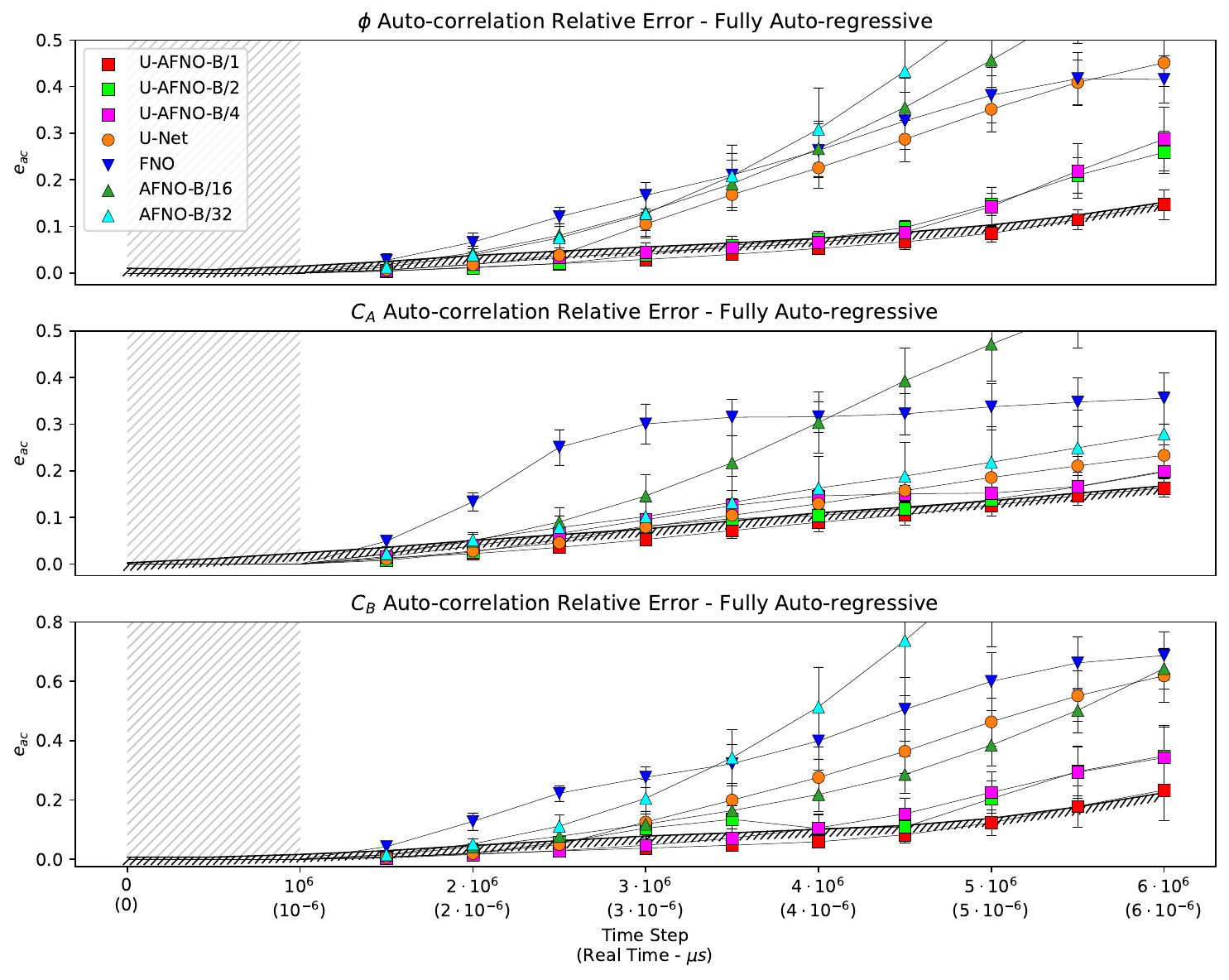}
    \caption{Auto-correlation relative errors for \textbf{fully auto-regressive} simulations. Each model is trained to skip \num{50000} time steps at-a-time, but for simplicity we only show the error values every \num{500000} steps. The hatched grey regions represent the initial $10^6$ time steps computed with the high fidelity solver. The cross-hatched black lines in each plot represent the average auto-correlation discrepancy between each distinct pair of ground truth simulations in the test set.}
    \label{fig:eac_far}
\end{figure}

The auto-correlation relative error metrics for each surrogate model are represented in Figure~\ref{fig:eac_far}. The auto-correlation error for the solid phase, and species A and B are shown every \num{500000} time steps. For each field, the U-AFNO-B/1 clearly outperforms all the baseline models. At the end of the simulation, the relative error for $\phi$ and $c_\text{A}$ does not exceed $15\%$,  while that for $c_\text{B}$ does not exceed $20\%$. Hence, even after 100 forward pass through the U-AFNO model, the error remains remarkably well controlled. When employing a larger patch size ($2\times2$ and $4\times4$), the errors slightly deteriorate, which is expected since larger transformer patch sizes generally yield worse accuracy. Yet, each U-AFNO model outperforms the other baseline models. In particular, they are more accurate than both U-Nets and vanilla AFNOs, confirming that combining these two model architectures can significantly improve prediction accuracy. The U-AFNO-B/1 relative error is on-par with the average discrepancy between distinct pairs of ground truth simulation (shown in hatched black). This is significant: in practice, it means that, on the auto-correlation relative error metric, and considering a given high-fidelity simulation, our surrogate will reproduce the dynamics equally well as any other high-fidelity run (on perturbatively different initial condition). 

\begin{figure}[!ht]
\centering
\includegraphics[width=1\textwidth]{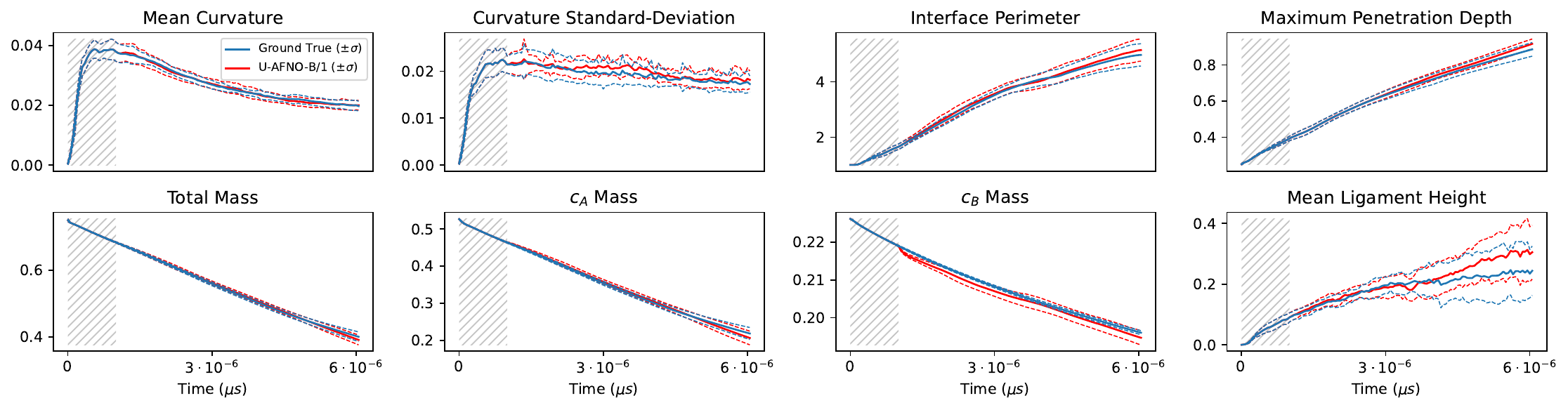}
    \caption{QoI dynamics predicted with the U-AFNO-B/1 (fully auto-regressive case). For each QoI, the solid blue line is the mean QoI value over time, taken across all the ground truth simulations in the test set (and the dotted lines are the corresponding standard deviations). The solid red lines represent the QoI dynamics predicted by the U-AFNO (averaged over all test surrogate simulations).}
    \label{fig:qoifar_uafno1}
\end{figure}

\begin{figure}[!ht]
\centering
\includegraphics[width=1\textwidth]{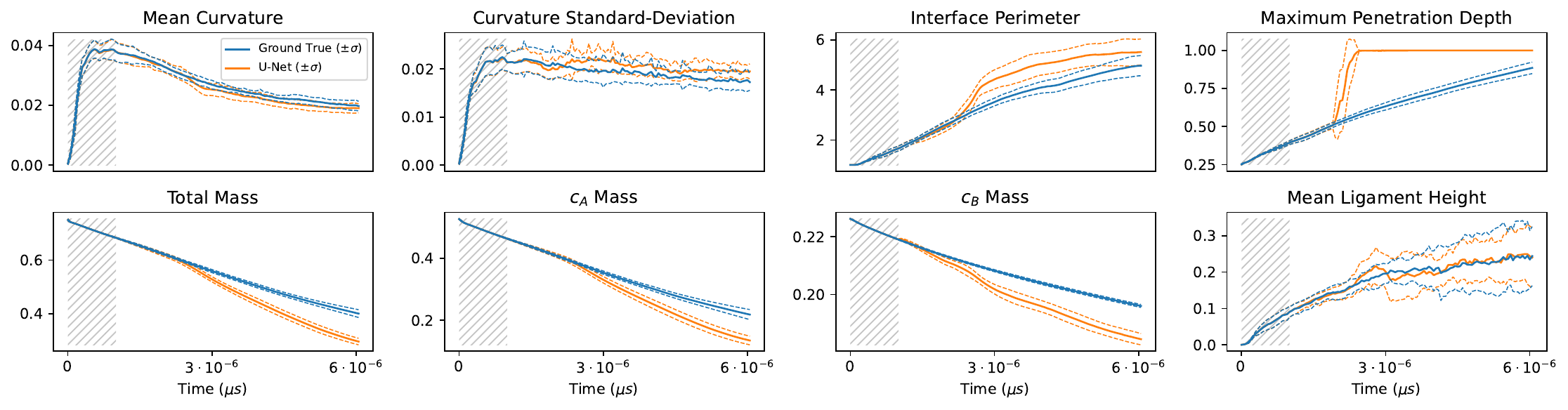}
    \caption{QoI dynamics predicted with the U-Net (fully auto-regressive case). For each QoI, the solid blue line is the mean QoI value over time, taken across all the ground truth simulations in the test set (and the dotted lines are the corresponding standard deviations). The solid red lines represent the QoI dynamics predicted by the U-Net (averaged over all test surrogate simulations).}
    \label{fig:qoifar_unet}
\end{figure}

The QoI relative errors are provided in Table~\ref{tab:eac_far}, and the predicted dynamics of each QoI are shown in Figure~\ref{fig:qoifar_uafno1} and Figure~\ref{fig:qoifar_unet} for the U-AFNO-B/1 and the U-Net, respectively. Note that the QoIs and error metrics are all computed across 20 test simulations, therefore the values in Table~\ref{tab:eac_far} and curves in Figure~\ref{fig:qoifar_uafno1} and \ref{fig:qoifar_unet} are mean values (with standard deviations). The U-AFNO clearly outperforms the baseline models and predicts the evolution of each QoI with very reasonable accuracy, as it rarely exceeds $5$ to $10\%$ relative error (except for the mean ligament height, with a mean relative error of $36.5\%$). As shown in Figure~\ref{fig:qoifar_uafno1}, our proposed model reproduces the mean curvature and mass envelopes with remarkable accuracy, and is also able to capture the curvature standard-deviation, interface perimeter and maximum penetration depth true patterns very well. Conversely, in Figure~\ref{fig:qoifar_unet}, the U-Net underestimates the mean curvature and mass envelope values, while overestimating the curvature standard-deviation, interface perimeter and maximum penetration depth. Most notably, with the U-Net, the liquid penetrates into the solid exceedingly fast and reaches the bottom boundary much earlier then it should (the maximum penetration depth steeply rises after 2~$\mu$s). This is due to the fact that with the U-Net, the liquid-solid interface tends to collapse at both the right and left edges of the domain. It is not clear why this phenomena occurs, but a possible explanation is that the U-Net fails to accurately capture the effects of the boundary conditions on the field dynamics.

\begin{table}[!ht]
\centering
\resizebox{\linewidth}{!}{
\begin{tabular}{lcccccccc}
\Xhline{3\arrayrulewidth}\\[-6pt]
Model & \makecell{Mean\\Curvature} & \makecell{Curvature\\ Standard Deviation} & \makecell{Interface\\ Perimeter} & Total Mass & $c_A$ Mass & $c_B$ Mass & \makecell{Maximum\\Penetration Depth} & \makecell{Mean Ligament\\ Height}\\[8pt]\hline\\[-6pt]
U-AFNO-B/1 (Ours) & $\pmb{0.057\pm0.019}$ & $\pmb{0.117\pm0.032}$ & $\pmb{0.068\pm0.029}$ & $\pmb{0.011\pm0.006}$ & $\pmb{0.023\pm0.011}$ & $\pmb{0.007\pm0.003}$ & $\pmb{0.040\pm0.017}$ & $0.358\pm0.238$\\
U-AFNO-B/2 (Ours) & $0.082\pm0.030$ & $0.128\pm0.044$ & $0.182\pm0.038$ & $0.032\pm0.009$ & $0.057\pm0.017$ & $0.014\pm0.004$ & $0.055\pm0.023$ & $0.345\pm0.096$\\
U-AFNO-B/4 (Ours) & $0.058\pm0.017$ & $0.122\pm0.030$ & $0.073\pm0.033$ & $0.036\pm0.012$ & $0.055\pm0.019$ & $0.029\pm0.004$ & $0.052\pm0.022$ & $\pmb{0.328\pm0.150}$\\
U-Net & $0.069\pm0.022$ & $0.144\pm0.043$ & $0.202\pm0.076$ & $0.096\pm0.015$ & $0.119\pm0.026$ & $0.033\pm0.006$ & $0.364\pm0.037$ & $0.368\pm0.165$\\
FNO & $1.862\pm0.290$ & $3.798\pm0.310$ & $0.282\pm0.073$ & $0.107\pm0.021$ & $0.164\pm0.031$ & $0.054\pm0.019$ & $0.326\pm0.107$ & $0.841\pm0.047$\\
AFNO-B/16 & $0.790\pm0.092$ & $1.895\pm0.173$ & $0.132\pm0.053$ & $0.123\pm0.026$ & $0.106\pm0.022$ & $0.133\pm0.036$ & $0.040\pm0.016$ & $0.366\pm0.146$\\
AFNO-B/32 & $0.777\pm0.081$ & $2.110\pm0.269$ & $0.286\pm0.085$ & $0.121\pm0.035$ & $0.211\pm0.049$ & $0.045\pm0.032$ & $0.123\pm0.059$ & $0.581\pm0.109$\\
[2pt]\Xhline{3\arrayrulewidth}
\end{tabular}}
\vspace{0.05in}
\caption{QoIs relative errors with \textbf{fully auto-regressive} roll-out. For U-AFNOs and vanilla AFNOs, B/1, B/2 and B/4 indicate a patch size of $1\times1$, $2\times2$ and $4\times4$, respectively. U-AFNOs can employ significantly smaller patch sizes than AFNOs, since the U-Net encoder architecture downsamples the AFNO input. Note that for fair comparison, every model output is wrapped into the interval $[0, 1]$, as described in section~\ref{sec:uafno-subsec}}
\label{tab:eac_far}
\end{table}

\begin{figure}[!ht]
\centering
    \includegraphics[width=0.85\textwidth]{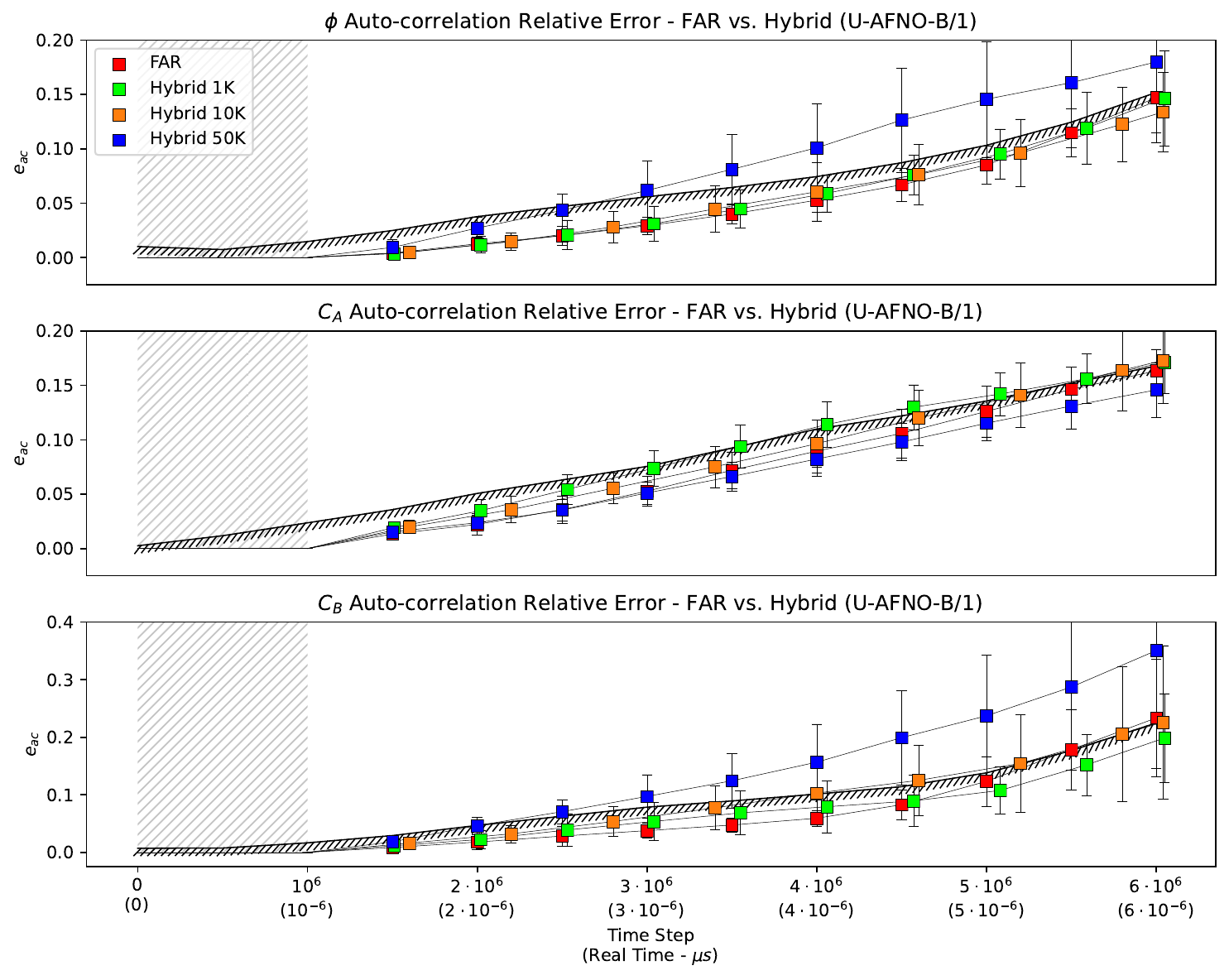}
    \caption{Auto-correlation relative errors for the U-AFNO-B/1, using either fully auto-regressive simulations, or hybrid simulations with 1000, \num{10000}, and \num{50000} relaxation time steps. The cross-hatched black lines in each plot represent the average auto-correlation discrepancy between each distinct pair of ground truth simulations in the test set. Note that when using either 1000 or \num{10000} relaxation steps, the timeline of the available snapshot is progressively shifting. For example, in the fully auto-regressive case (or the \num{50000} relaxation steps case), the \num{1050000}\textsuperscript{th} step and the \num{1100000}\textsuperscript{th} step will be consecutively available. Conversely, with \num{10000} relaxation steps, the \num{1050000}\textsuperscript{th}, \num{1060000}\textsuperscript{th} and \num{1110000}\textsuperscript{th} steps will be consecutively available. This is the reason why the 1000 relaxation steps (green) and \num{10000} relaxation step (orange) markers are slightly shifted.}
    \label{fig:eac_far_vs_hyb_uafno}
\end{figure}

\begin{figure}[!ht]
\centering
    \includegraphics[width=0.85\textwidth]{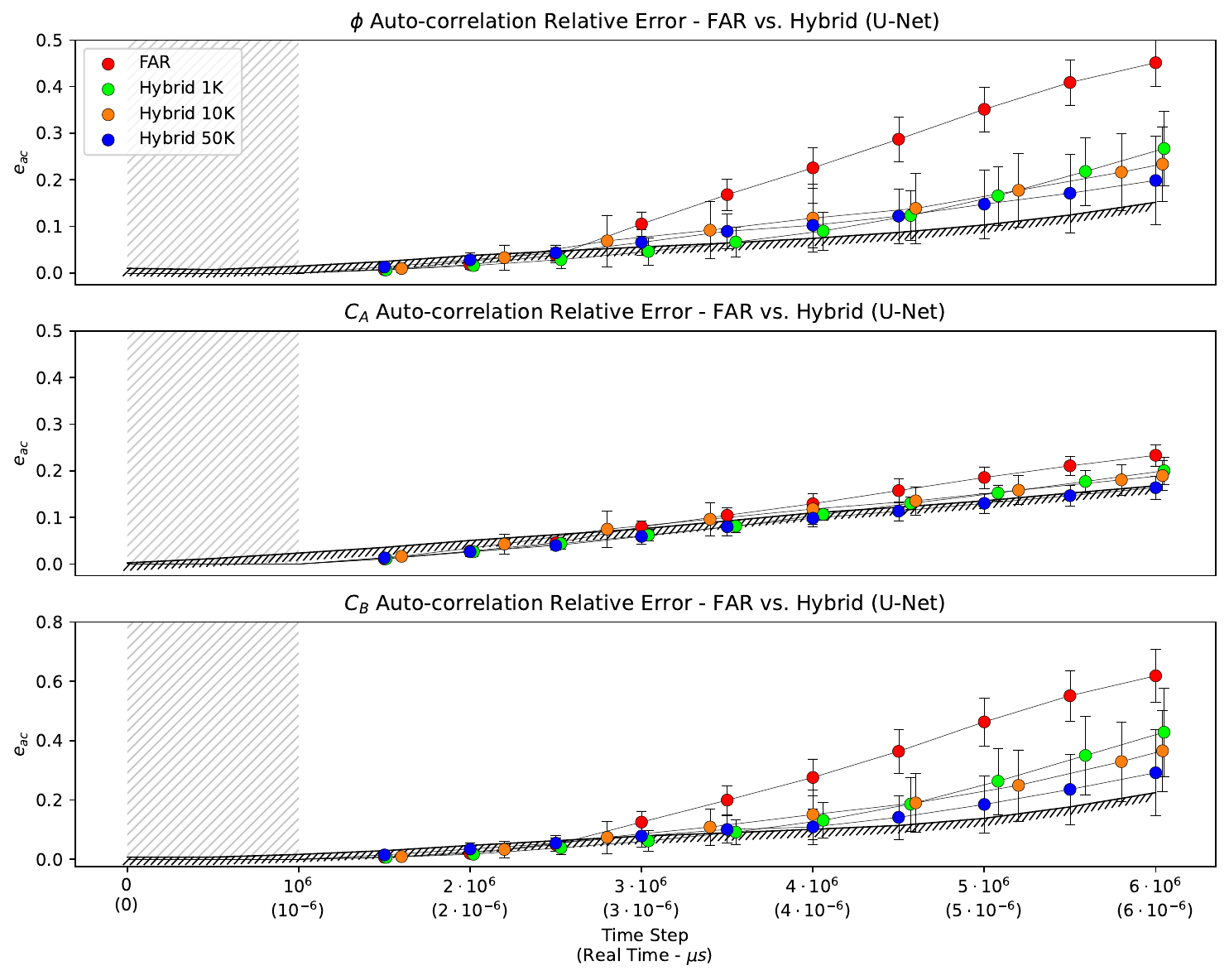}
    \caption{Auto-correlation relative errors for the U-Net, using either fully auto-regressive simulations, or hybrid simulations with 1,000, 10,000, and 50,000 relaxation time steps. The cross-hatched black lines in each plot represent the average auto-correlation discrepancy between each distinct pair of ground truth simulations in the test set.}
    \label{fig:eac_far_vs_hyb_unet}
\end{figure}

We now investigate the effects of adding hybrid time stepping, as briefly described in section \ref{sec:hybrid}. We use the same trained models as in the fully auto-regressive case (leaping \num{50000} time steps at-a-time), but after each forward pass through the surrogate model, we now employ the high-fidelity solver to advance in time for a limited number of relaxation time steps. For both the U-AFNO and the baseline models, we run test simulations for \num{1000}, \num{10000} and \num{50000} relaxation time steps. Figure~\ref{fig:eac_far_vs_hyb_uafno} and Figure~\ref{fig:eac_far_vs_hyb_unet} show the auto-correlation relative errors with and without relaxation steps, for the U-AFNO and the U-Net, respectively. With the U-AFNO, the accuracy of the surrogate simulation does not significantly improve when using high-fidelity relaxation time stepping. While this may seem surprising, it is in fact reasonable since the margin for improvement is limited. Indeed, as seen in the fully auto-regressive case, the U-AFNO average error is already on par with the average auto-correlation discrepancy between distinct ground truth simulations. Thus, reducing the error further with a model that still relies on surrogate modeling (whether moderately or not) is unlikely to succeed. 

For the U-Net however, we observe a different trend. As seen in Figure~\ref{fig:eac_far_vs_hyb_unet}, the error significantly decreases when employing high-fidelity relaxation steps, likely because there is more room for improvement (compared to the fully auto-regressive case). With just 1,000 relaxation time steps, the errors sharply decrease (especially for $\phi$ and $c_B$), indicating that even just a few relaxation time steps are sufficient to correct the field physics. When increasing the number of relaxation time steps, the errors eventually decrease further, but with diminishing returns. This is expected since (as discussed in the previous paragraph), it is unlikely for any surrogate model to achieve significantly lower errors than the average auto-correlation discrepancy between ground truth simulations. The observation that blending U-Net forward passes with high-fidelity time stepping to improve accuracy is very consistent with findings in Oommen \textit{et al.}~\cite{oommen2023rethinking}. However, as seen with our proposed U-AFNO model and in Figure~\ref{fig:eac_far_vs_hyb_uafno}, the gains from employing high-fidelity time stepping may vary depending on the specific surrogate model architecture. 

Tables~\ref{tab:eac_hyb1k}, \ref{tab:eac_hyb10k}, and \ref{tab:eac_hyb50k} show the QoI relative errors for \num{1000}, \num{10000} and \num{50000} high-fidelity relaxation time steps, respectively. While the U-AFNO still generally outperforms the baseline models (in particular with a patch size of 1$\times$1), the improvements compared to the fully auto-regressive case are marginal, and sometimes the accuracy even degrades (as already observed with the auto-correlation errors in Figure~\ref{fig:eac_far_vs_hyb_uafno}). For example, the relative error for the mass of species B increases from $0.7\%$ in the fully auto-regressive case to $3.2\%$ in the hybrid (1000 relaxation steps) case. With some of the baseline models, such as the FNO and the vanilla AFNOs, the accuracy slightly improves between the fully auto-regressive case and the hybrid case (with 1000 relaxation steps), but then significantly deteriorates with \num{10000} and \num{50000} relaxation steps. In these latter two cases, the metal tends to be consumed by the corrosive liquid much faster than it normally should, and fully disappears early-on. As a result, the liquid-metal interface is not properly defined anymore, and the QoIs depending on this interface are unavailable (e.g. curvature QoIs, etc.). Intuitively, the accuracy should improve when employing more and more relaxation time steps, as a larger and larger portion of the simulation is computed through the high-fidelity solver. However as discussed earlier, this assertion does not always hold true. Except for the vanilla U-Net, the accuracy either remains stable, or worsens. Additional investigation as to why and when accuracy may degrade is left to future work, but we conjecture that this degradation is linked to the robustness of the high-fidelity solver to un-physical initial conditions. Since the neural operator is purely data-driven and no physical constraints are enforced, the neural operator outputs may violate some underlying assumptions of the high-fidelity solver, leading to instability or un-physical behavior build-up by the high-fidelity solver itself. For instance, in the high-fidelity LMD simulations, periodic boundary conditions are enforced between the left and right edges of the domain. However, these boundary conditions are not hard-enforced during the neural operator training. Thus, the neural operator outputs, which serve as initial conditions for the high-fidelity solver may not be strictly periodic, leading to unexpected behaviors.   

The high-fidelity solver average wall clock run time is 0.026$s$ per time step (on 128 CPU cores). That is, 1303$s$ per \num{50000} steps. The average wall clock run-time for a forward pass through the U-AFNO (with a $1\times1$ patch size) is 0.116$s$ on a single GPU. Thus, the speed-up to compute \num{50000} time steps is about \num{11200}$\times$. The U-AFNO run time is quasi-negligible compared to the high-fidelity solver run time, so the speed-up of a full surrogate simulation (in the fully auto-regressive case) comes down to the time taken to initialize the simulation. In this paper, we have run 6$\mu$s worth of real-time simulation, where the initial 1$\mu$s is run with the high-fidelity solver (i.e. $10^6$ time steps), and the remaining 5$\mu$s are obtained with U-AFNO auto-regressive roll-out. This is equivalent to a 6$\times$ speed-up. Note that augmenting the surrogate simulation with high-fidelity relaxation steps would considerably lower the speed up gains, and as discussed in the previous paragraph, the U-AFNO already achieve excellent accuracy without needing such hybrid time stepping. 

\begin{table}[!ht]
\centering
\resizebox{\linewidth}{!}{
\begin{tabular}{lcccccccc}
\Xhline{3\arrayrulewidth}\\[-6pt]
Model & \makecell{Mean\\Curvature} & \makecell{Curvature\\ Standard Deviation} & \makecell{Interface\\ Perimeter} & Total Mass & $c_A$ Mass & $c_B$ Mass & \makecell{Maximum\\Penetration Depth} & \makecell{Mean Ligament\\ Height}\\[8pt]\hline\\[-6pt]
U-AFNO-B/1 (Ours) & $\pmb{0.054\pm0.017}$ & $\pmb{0.114\pm0.035}$ & $\pmb{0.083\pm0.045}$ & $\pmb{0.017\pm0.009}$ & $\pmb{0.028\pm0.012}$ & $0.032\pm0.006$ & $\pmb{0.031\pm0.019}$ & $\pmb{0.317\pm0.221}$\\
U-AFNO-B/2 (Ours) & $0.056\pm0.017$ & $0.138\pm0.042$ & $0.139\pm0.047$ & $0.036\pm0.011$ & $0.060\pm0.019$ & $0.015\pm0.003$ & $0.059\pm0.024$ & $0.355\pm0.220$\\
U-AFNO-B/4 (Ours) & $0.060\pm0.017$ & $0.117\pm0.037$ & $0.105\pm0.063$ & $0.019\pm0.013$ & $0.046\pm0.021$ & $0.030\pm0.005$ & $0.036\pm0.021$ & $0.350\pm0.201$\\
U-Net & $0.066\pm0.024$ & $0.145\pm0.052$ & $0.129\pm0.062$ & $0.040\pm0.021$ & $0.067\pm0.032$ & $\pmb{0.012\pm0.006}$ & $0.104\pm0.098$ & $0.381\pm0.202$\\
FNO & $0.735\pm0.077$ & $1.961\pm0.215$ & $0.342\pm0.053$ & $0.342\pm0.062$ & $0.375\pm0.060$ & $0.320\pm0.059$ & $0.266\pm0.104$ & $0.759\pm0.073$\\
AFNO-B/16 & $0.137\pm0.039$ & $0.471\pm0.105$ & $0.406\pm0.051$ & $0.250\pm0.021$ & $0.185\pm0.033$ & $0.336\pm0.019$ & $0.054\pm0.038$ & $0.433\pm0.167$\\
AFNO-B/32 & $0.202\pm0.034$ & $0.651\pm0.113$ & $0.357\pm0.080$ & $0.268\pm0.057$ & $0.229\pm0.051$ & $0.329\pm0.066$ & $0.096\pm0.041$ & $0.400\pm0.123$\\
[2pt]\Xhline{3\arrayrulewidth}
\end{tabular}}
\vspace{0.05in}
\caption{QoIs relative errors with a hybrid surrogate/high-fidelity roll-out. Here, \textbf{1000 high-fidelity} relaxation time steps are employed after each forward pass through the surrogate model. In some cases, the surrogate simulation is too inaccurate to exhibit a properly defined liquid-metal interface, in which case any QoI depending on this interface is unavailable (listed as N/A).}
\label{tab:eac_hyb1k}
\end{table}

\begin{table}[!ht]
\centering
\resizebox{\linewidth}{!}{
\begin{tabular}{lcccccccc}
\Xhline{3\arrayrulewidth}\\[-6pt]
Model & \makecell{Mean\\Curvature} & \makecell{Curvature\\ Standard Deviation} & \makecell{Interface\\ Perimeter} & Total Mass & $c_A$ Mass & $c_B$ Mass & \makecell{Maximum\\Penetration Depth} & \makecell{Mean Ligament\\ Height}\\[8pt]\hline\\[-6pt]
U-AFNO-B/1 (Ours) & $\pmb{0.056\pm0.016}$ & $\pmb{0.114\pm0.038}$ & $\pmb{0.084\pm0.045}$ & $\pmb{0.013\pm0.007}$ & $0.035\pm0.015$ & $0.028\pm0.006$ & $0.042\pm0.025$ & $\pmb{0.261\pm0.129}$\\
U-AFNO-B/2 (Ours)  & $\pmb{0.056\pm0.018}$ & $0.116\pm0.033$ & $0.089\pm0.047$ & $0.023\pm0.010$ & $0.045\pm0.019$ & $\pmb{0.018\pm0.004}$ & $0.047\pm0.021$ & $0.270\pm0.194$\\
U-AFNO-B/4 (Ours)  & $0.057\pm0.015$ & $0.123\pm0.051$ & $0.110\pm0.054$ & $0.025\pm0.014$ & $\pmb{0.034\pm0.018}$ & $0.029\pm0.006$ & $\pmb{0.040\pm0.027}$ & $0.291\pm0.166$\\
U-Net  & $0.060\pm0.018$ & $0.125\pm0.037$ & $0.087\pm0.033$ & $0.044\pm0.030$ & $0.062\pm0.035$ & $0.021\pm0.015$ & $0.101\pm0.109$ & $0.336\pm0.165$\\
FNO  & N/A & N/A & N/A & $0.540\pm0.055$ & $0.540\pm0.040$ & $0.534\pm0.079$ & N/A & N/A\\
AFNO-B/16  & N/A & N/A & N/A & $0.580\pm0.124$ & $0.522\pm0.117$ & $0.652\pm0.127$ & N/A & N/A\\
AFNO-B/32  & N/A & N/A & N/A & $0.791\pm0.020$ & $0.703\pm0.024$ & $0.843\pm0.015$ & N/A & N/A\\
[2pt]\Xhline{3\arrayrulewidth}
\end{tabular}}
\vspace{0.05in}
\caption{QoIs relative errors with a hybrid surrogate/high-fidelity roll-out. Here, \textbf{10,000 high-fidelity} relaxation time steps are employed after each forward pass through the surrogate model. In some cases, the surrogate simulation is too inaccurate to exhibit a properly defined liquid-metal interface, in which case any QoI depending on this interface is unavailable (listed as N/A).}
\label{tab:eac_hyb10k}
\end{table}

\begin{table}[!ht]
\centering
\resizebox{\linewidth}{!}{
\begin{tabular}{lcccccccc}
\Xhline{3\arrayrulewidth}\\[-6pt]
Model & \makecell{Mean\\Curvature} & \makecell{Curvature\\ Standard Deviation} & \makecell{Interface\\ Perimeter} & Total Mass & $c_A$ Mass & $c_B$ Mass & \makecell{Maximum\\Penetration Depth} & \makecell{Mean Ligament\\ Height}\\[8pt]\hline\\[-6pt]
U-AFNO-B/1 (Ours) & $0.061\pm0.018$ & $\pmb{0.107\pm0.036}$ & $0.092\pm0.042$ & $\pmb{0.029\pm0.013}$ & $0.053\pm0.019$ & $\pmb{0.014\pm0.004}$ & $\pmb{0.058\pm0.028}$ & $0.272\pm0.132$\\
U-AFNO-B/2 (Ours) & $\pmb{0.051\pm0.013}$ & $0.108\pm0.037$ & $\pmb{0.076\pm0.032}$ & $0.034\pm0.011$ & $0.041\pm0.016$ & $0.021\pm0.004$ & $0.059\pm0.027$ & $0.271\pm0.151$\\
U-AFNO-B/4 (Ours) & $0.055\pm0.017$ & $\pmb{0.107\pm0.036}$ & $0.109\pm0.051$ & $0.041\pm0.012$ & $0.048\pm0.018$ & $0.018\pm0.005$ & $\pmb{0.058\pm0.024}$ & $\pmb{0.263\pm0.152}$\\
U-Net & $0.065\pm0.027$ & $0.110\pm0.039$ & $0.102\pm0.064$ & $0.030\pm0.019$ & $\pmb{0.039\pm0.023}$ & $0.017\pm0.006$ & $0.083\pm0.067$ & $0.266\pm0.101$\\
FNO & N/A & N/A & N/A & $0.400\pm0.076$ & $0.394\pm0.073$ & $0.414\pm0.075$ & N/A & N/A\\
AFNO-B/16 & N/A & N/A & N/A & $0.493\pm0.108$ & $0.443\pm0.124$ & $0.635\pm0.228$ & N/A & N/A\\
AFNO-B/32 & N/A & N/A & N/A & $0.734\pm0.027$ & $0.669\pm0.030$ & $0.799\pm0.020$ & N/A & N/A\\
[2pt]\Xhline{3\arrayrulewidth}
\end{tabular}}
\vspace{0.05in}
\caption{QoIs relative errors with a hybrid surrogate/high-fidelity roll-out. Here, \textbf{50,000 high-fidelity} relaxation time steps are employed after each forward pass through the surrogate model. In some cases, the surrogate simulation is too inaccurate to exhibit a properly defined liquid-metal interface, in which case any QoI depending on this interface is unavailable (listed as N/A).}
\label{tab:eac_hyb50k}
\end{table}


\FloatBarrier

\section{Conclusion}
\label{sec:concl}

In this paper, we have introduced U-AFNOs, a new ML-based surrogate model for fast prediction of time-dependant PDEs. This architecture was applied to phase field simulations, specifically the dealloying corrosion of metals, where the infiltration of a corrosive liquid into the alloy can rapidly lead to morphologically-complex metal structures. Employed within an auto-regressive roll-out, our proposed model is able to reproduce the correct LMD patterns, and accurately captures invariant statistics, even in high-chaotic simulations. Our work supports the idea that combining U-Nets and vision transformers may provide very promising architectures for operator learning. Furthermore, we have identified meaningful QoIs to accurately describe the LMD dynamics and showed that they can be accurately predicted by our model. We have also investigated the relevance of augmenting auto-regressive surrogate simulations with hybrid high-fidelity time stepping. We showed that the potential gains in accuracy are not systematic, and may be dependent on the architecture of the ML surrogate and/or the high-fidelity solver (i.e. problem-dependent). 

\section*{Acknowledgements}
This work was supported by the U.S. Department of Energy, Office of Nuclear Energy, and Office of Science, Office of Advanced Scientific Computing Research through the Scientific Discovery through Advanced Computing project on Simulation of the Response of Structural Metals in Molten Salt Environment.
This article has been co-authored by employees of National Technology and Engineering Solutions of Sandia, LLC under Contract No. DE-NA0003525 with the U.S. Department of Energy (DOE). The employees co-own right, title and interest in and to the article and are responsible for its contents. The United States Government retains and the publisher, by accepting the article for publication, acknowledges that the United States Government retains a non-exclusive, paid-up, irrevocable, world-wide license to publish or reproduce the published form of this article or allow others to do so, for United States Government purposes. The DOE will provide public access to these results of federally sponsored research in accordance with the DOE Public Access Plan \url{https://www.energy.gov/downloads/doe-public-access-plan}. Sandia Release Number: SAND2024-07895O.
Los Alamos National Laboratory, United States, an affirmative action/equal opportunity employer, is operated by Triad National Security, LLC, for the National Nuclear Security Administration of the U.S. Department of Energy under Contract No. 89233218CNA000001.
Lawrence Berkeley National Laboratory is supported by the DOE Office of Science under contract no. DE-AC02-05CH11231. This study made use of computational resources of the National Energy Research Scientific Computing Center (NERSC), which is also supported by the Office of Basic Energy Sciences of the US Department of Energy under the same contract number.
%

\bibliographystyle{unsrt}
\bibliography{references}

\end{document}